\RequirePackage[mathlines]{lineno}
\documentclass[amsmath,amssymb,twocolumn,prd,floatfix,showpacs, nofootinbib]{revtex4-1}
\usepackage{tabularx}
\usepackage{bm, color}
\usepackage{overpic,subfigure}
\usepackage{multirow}
\usepackage{array}
\usepackage{dcolumn}
\usepackage[symbol]{footmisc}
\usepackage{epstopdf}
\usepackage{lineno}
\usepackage{hyperref}
\usepackage{amsmath}

\hypersetup{hypertex=true,
	        colorlinks = true,
            linkcolor = blue,
            urlcolor = blue,
            citecolor = blue}
\newcommand{\BESIIIorcid}[1]{\href{https://orcid.org/#1}{\hspace*{0.1em}\raisebox{-0.45ex}{\includegraphics[width=1em]{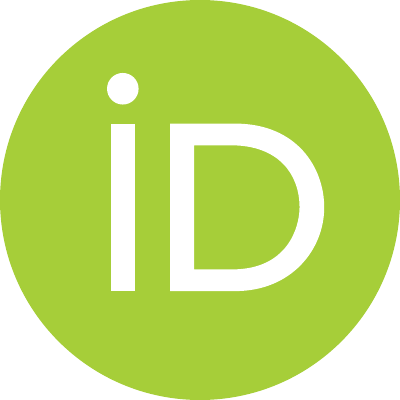}}}}
\begin{document}
\title{\boldmath Observation of $\chi_{c1,2}\to p K^- K^- \bar \Xi^{+} +c.c.$ and search for $\chi_{c0}\to p K^- K^- \bar \Xi^{+} +c.c.$ }
%% Saved at => 2026-04-29
\author{
M.~Ablikim$^{1}$\BESIIIorcid{0000-0002-3935-619X},
M.~N.~Achasov$^{4,c}$\BESIIIorcid{0000-0002-9400-8622},
P.~Adlarson$^{84}$\BESIIIorcid{0000-0001-6280-3851},
X.~C.~Ai$^{90}$\BESIIIorcid{0000-0003-3856-2415},
C.~S.~Akondi$^{32A,32B}$\BESIIIorcid{0000-0001-6303-5217},
R.~Aliberti$^{40}$\BESIIIorcid{0000-0003-3500-4012},
A.~Amoroso$^{83A,83C}$\BESIIIorcid{0000-0002-3095-8610},
Q.~An$^{79,66,\dagger}$,
Y.~H.~An$^{90}$\BESIIIorcid{0009-0008-3419-0849},
M.~S.~Anderson$^{40}$\BESIIIorcid{0009-0008-1550-2632}
% ...
\\
\vspace{0.2cm}
(BESIII Collaboration)\\
\vspace{0.2cm} {\it
$^{1}$ Institute of High Energy Physics, Beijing 100049, People's Republic of China\\
% ...
$^{4}$ Budker Institute of Nuclear Physics SB RAS (BINP), Novosibirsk 630090, Russia\\
% ...
$^{32}$ INFN Sezione di Ferrara, (A)INFN Sezione di Ferrara, I-44122, Ferrara, Italy; (B)University of Ferrara, I-44122, Ferrara, Italy\\
% ...
$^{40}$ Johannes Gutenberg University of Mainz, Johann-Joachim-Becher-Weg 45, D-55099 Mainz, Germany\\
% ...
$^{66}$ State Key Laboratory of Particle Detection and Electronics, Beijing 100049, Hefei 230026, People's Republic of China\\
% ...
$^{79}$ University of Science and Technology of China, Hefei 230026, People's Republic of China\\
% ...
$^{83}$ University of Turin and INFN, (A)University of Turin, I-10125, Turin, Italy; (B)University of Eastern Piedmont, I-15121, Alessandria, Italy; (C)INFN, I-10125, Turin, Italy\\
$^{84}$ Uppsala University, Box 516, SE-75120 Uppsala, Sweden\\
% ...
$^{90}$ Zhengzhou University, Zhengzhou 450001, People's Republic of China\\
\vspace{0.2cm}
$^{\dagger}$ Deceased\\
% ...
$^{c}$ Also at the Novosibirsk State University, Novosibirsk, 630090, Russia\\
% ...
}
}
%% ends here %%

\begin{abstract}
  By analyzing $(2.712\pm0.014)\times10^9$ $\psi(3686)$ events
  collected by the BESIII detector, we report the first experimental
  study of the decays
  $\chi_{cJ}\to p K^- K^- \bar \Xi^{+}+c.c.~(J=0,1,2)$.  The decays
  $\chi_{c1}\to p K^- K^- \bar \Xi^{+}+c.c.$ and
  $\chi_{c2}\to p K^- K^- \bar \Xi^{+}+c.c.$ are observed for
  the first time with significances of $9.0\sigma$ each, and their branching fractions are determined to be
  $\mathcal{B}_{\chi_{c1}\to p K^- K^- \bar \Xi^{+}+c.c.} =(9.6^{+2.2}_{-2.0}\pm0.8)\times10^{-6}$, and
  $\mathcal{B}_{\chi_{c2}\to p K^- K^- \bar \Xi^{+}+c.c.} =(7.5^{+1.8}_{-1.5}\pm0.5)\times10^{-6}$, where the first
  and second uncertainties are statistical and systematic,
  respectively.  No significant signal is observed for
  $\chi_{c0}\to p K^- K^- \bar \Xi^{+}+c.c.$, and an upper
  limit on its branching fraction of $12.7\times10^{-6}$ is established at
  the 90\% confidence level.
\end{abstract}

\maketitle

\section{Introduction}
Charmonium decays serve as essential probes for testing perturbative quantum chromodynamics~(QCD) predictions and constraining non-perturbative QCD models~\cite{pqcd1,pqcd2,pqcd3}.
The $\chi_{cJ}$ states $(J=0,1,2)$, representing $P$-wave charmonium configurations~\cite{ref::abc,ref::abcd}, are particularly important in this context. Compared to the well-studied $J/\psi$ and $\psi(3686)$ meson decays,
the understanding of $\chi_{cJ}$ decays remains limited. Many decay channels of these charmonium states have not been experimentally studied and their branching fractions~(BFs) undetermined.

Recently, several three-body $\psi(3686)$ and $\chi_{cJ}$ decays involving three strange particles have been reported, including $K^- \Lambda \bar \Xi^+$~\cite{dd1}, $K^- \Sigma^0  \bar \Xi^+$~\cite{dd2}, $\Lambda(1520)K^-\bar \Xi^+$~\cite{dd3},
$\Xi^- K_S^0 \bar \Omega^+$~\cite{dd4} and $\Omega^- K^+ \bar \Xi^0$~\cite{dd5}. However, no observations of $\chi_{cJ}$ decays into final states with three strange particles in the $\chi_{cJ}\to BB'KK$ system have been reported,
where $B$ denotes a strange baryon and $B'$ a light baryon. Such decays, involving the production of multiple strange quark pairs, provide unique insights into the mechanisms of multi-strange hadron formation and the dynamics of strange quark hadronization.

In this paper, we report a search for the four-body decay $\chi_{cJ} \to p K^- K^- \bar \Xi^{+}$ using $(2.712\pm0.014)\times10^9$ $\psi(3686)$ events collected with the BESIII detector~\cite{psip-num-inc}. Charge-conjugate modes are implicitly included throughout.

\section{BESIII DETECTOR AND MONTE CARLO SIMULATION}
\label{sec:BES}
The BESIII detector~\cite{ref::detector} records symmetric $e^+e^-$ collisions provided by the BEPCII storage ring~\cite{ref::collider}
in the center-of-mass energy range from 1.84 to 4.95~GeV, with a peak luminosity of $1.1 \times 10^{33}\;\text{cm}^{-2}\text{s}^{-1}$ at $\sqrt{s} = 3.773\;\text{GeV}$.

The cylindrical core of the BESIII detector covers 93\% of the full solid angle and consists of a helium-based multilayer drift chamber~(MDC),
a plastic scintillator time-of-flight system~(TOF), and a CsI(Tl) electromagnetic calorimeter~(EMC), which are all enclosed in a superconducting
solenoidal magnet providing a 1.0~T magnetic field. The solenoid is supported by an octagonal flux-return yoke with resistive plate counter muon
identification modules interleaved with steel. The charged-particle momentum resolution at $1~{\rm GeV}/c$ is $0.5\%$, and the ${\rm d}E/{\rm d}x$
resolution is $6\%$ for electrons from Bhabha scattering. The EMC measures photon energies with a resolution of $2.5\%$ ($5\%$) at $1$~GeV in the barrel (end-cap) region. The time resolution in the TOF barrel region is 68~ps, while that in the end-cap region was 110~ps.
The end-cap TOF system was upgraded in 2015 using multi-gap resistive plate chamber technology, achieving a time resolution of 60~ps~\cite{Tof1,Tof2,Tof3}.
Around $83\%$ of the data used in this analysis benefits from this upgrade.

Simulated Monte Carlo (MC) samples are produced using a {\sc
geant4}-based~\cite{Geant4} MC package, which
includes the geometric description of the BESIII detector and its response. These samples are used to determine detection efficiencies
and to estimate backgrounds. The simulation models the beam
energy spread and initial state radiation~(ISR) in $e^+e^-$
annihilations with the generator {\sc
kkmc}~\cite{Jadach01}.

The inclusive MC sample includes the production of the
$\psi(3686)$ resonance, the ISR production of the $J/\psi$, and
the continuum processes incorporated in {\sc
kkmc}~\cite{Jadach01}.
All particle decays are modeled with {\sc
evtgen}~\cite{Lange01,Lange02} using BFs
either taken from the Particle Data Group~(PDG)~\cite{ref::pdg2024}, where available,
or otherwise estimated with {\sc lundcharm}~\cite{Lundcharm00, Lundcharm01}.
Final state radiation
from charged final state particles is incorporated using the {\sc
photos} package~\cite{PHOTOS}.
An inclusive MC sample of $2.7\times10^{9}$ generic $\psi(3686)$ decays is used for background studies.
The signal MC samples for $\psi(3686)\to\gamma\chi_{cJ}$ are generated using~\cite{Lange01,Lange02}, where the radiative transition modeled according to P2GC$J$ models. For each $\chi_{c0,1,2}$ states, $0.5\times10^6$ signal MC events are generated to determine the detection efficiency.
The cascade decays $\chi_{cJ}\to p K^- K^- \bar{\Xi}^+ $, $\bar{\Xi}^{+} \to \bar \Lambda \pi^{+}$, and $\bar \Lambda \to \bar p \pi^+$ are generated assuming a unifrom phase space distribution.

\section{EVENT SELECTION}
\label{sec:selection}

Candidate events are reconstructed from the charmonium
transition $\psi(3686)\to\gamma\chi_{cJ}$, followed by the hadronic
decays $\chi_{cJ}\to p K^- K^- \bar\Xi^{+}$,
$\bar\Xi^{+}\to\bar\Lambda\pi^{+}$, and
$\bar\Lambda\to\bar p\pi^{+}$.
The final state contains one photon from the radiative transition
and six charged tracks: one proton, two kaons, one pion from
the $\bar\Xi^{+}$ decay, and a $\bar p\pi^{+}$ pair from the
$\bar\Lambda$ decay. Therefore, signal events are required to
contain at least one photon candidate and at least six charged tracks.

Photon candidates are identified from electromagnetic showers reconstructed in the EMC, with a minimum deposited energy of 25~MeV in the barrel region
($|\cos \theta|< 0.80$, where $\theta$ is the polar angle with respect to the $z$-axis, the symmetry axis of the MDC) and 50~MeV in the end-cap region ($0.86 <|\cos \theta|< 0.92$).
Showers associated with charged tracks are rejected by requiring the open angle between the shower and the nearest charged track, extrapolated to the EMC surface from the interaction point~(IP), to exceed 10 degrees.
The EMC shower time is required to be within [0,700]\,ns of the event start time to suppress electronic noise and beam-unrelated backgrounds.

Charged tracks not originating from $\bar \Lambda$ decays are required to satisfy $|\rm{cos\theta}|<0.93$, $|V_{z}|<10$~cm, and $|V_{xy}|<1$~cm, where $|V_{z}|$ and $|V_{xy}|$ are the distances of the closest approach to the IP along $z$-axis and in the transverse plane, respectively.
Particle identification~(PID) combines d$E$/d$x$ measurements with TOF information to compute likelihood values $\mathcal{L}(h)$ for each hadron hypothesis, $h\in\{p,\pi,K\}$.
Tracks satisfying $\mathcal{L}(p)>\mathcal{L}(\pi)$ and $\mathcal{L}(p)>\mathcal{L}(K)$ are identified as protons,
while those satisfying $\mathcal{L}(K)>\mathcal{L}(\pi)$ and $\mathcal{L}(K)>\mathcal{L}(p)$ are identified as kaon candidates.

To reconstruct $\bar{\Lambda}$, all possible oppositely charged
track pairs are considered as $\bar p \pi^+$ candidates without
applying PID requirements. Mis-combinations are rejected by the
subsequent selection criteria. A vertex fit is performed on all
$\bar p \pi^+$ combinations, followed by a secondary vertex fit
on the surviving combinations to suppress mis-combinations,
requiring the decay length significance to exceed two. If
multiple candidates survive the selection in a single event,
only the combination with the smallest $\chi^2$ is retained.
The $\bar p \pi^+$ invariant mass is required to satisfy
$1.109$~GeV/$c^2$ $<M_{\bar p \pi^+}<1.123$~GeV/$c^2$,
corresponding to a $\pm 3\sigma$ window around the nominal
$\bar \Lambda$ mass.

\begin{figure}[htbp]
    \centering
	 \hspace{-0.33cm}
     \includegraphics[width=1.028\linewidth]{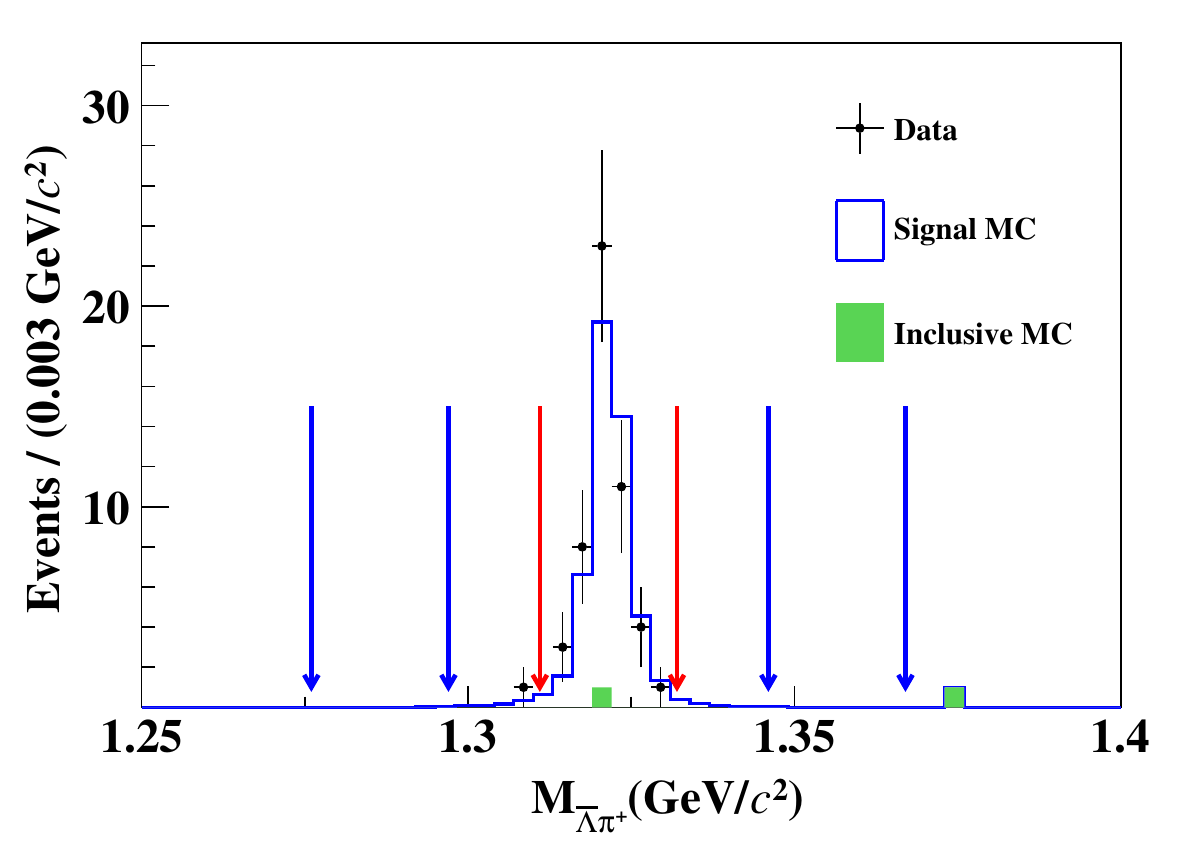}
	\vspace{-0.55cm}
    \caption{The distribution of $M_{\bar{\Lambda}\pi^+}$ for accepted candidate events. The dots with error bars are data, the green histogram represents the inclusive MC sample, and the blue histogram is the signal MC sample.
		The red and blue arrows indicate the signal and sideband regions, respectively.
    }
    \label{fig:AB}
\end{figure}

The $\bar \Xi^+$ candidate is reconstructed via the $\bar \Xi^+ \to \bar \Lambda \pi^+$ decay, where the
$\bar \Lambda$ candidate is required to satisfy the selection criteria described above.
The $\bar \Xi^+$ signal region is defined as $1.311$~GeV/$c^2$ $< M_{\bar {\Lambda} \pi^+} < 1.332$~GeV/$c^2$, corresponding to $\pm 3\sigma$ around the nominal $\bar \Xi^+$ mass.
The sideband regions are defined as $|M_{\bar{\Lambda}\pi^+}-1.2865|<0.0105~\mathrm{GeV}/c^2$ and $|M_{\bar{\Lambda}\pi^+}-1.3565|<0.0105~\mathrm{GeV}/c^2$, as shown in Fig.~\ref{fig:AB}.

A four-constraint (4C) kinematic fit is applied to each event under the hypothesis of $e^+ e^- \to \gamma p K^- K^- \bar \Lambda \pi^+ $,
enforcing conservation of energy and momentum.
If there is  multiple combinations in a single event, the one with the smallest  $\chi_{\rm 4C}^{2}$ is retained. The requirement $\chi^2_{\rm 4C}<50$ is imposed, optimized using the figure-of-merit $\mathit{S}/\sqrt{\mathit{S}+\mathit{B}}$, where $S$ is the number of signal events estimated from the signal MC sample scaled to the measured BF, and $B$ is the number of background events obtained from the inclusive $\psi(3686)$ MC sample. The continuum background, evaluated using control sample at $\sqrt s =3.650$~GeV~\cite{lum}, is found to be negligible.

\begin{figure}[htbp]
	\centering
	\hspace{-0.331cm}
	\includegraphics[width=1.03\linewidth]{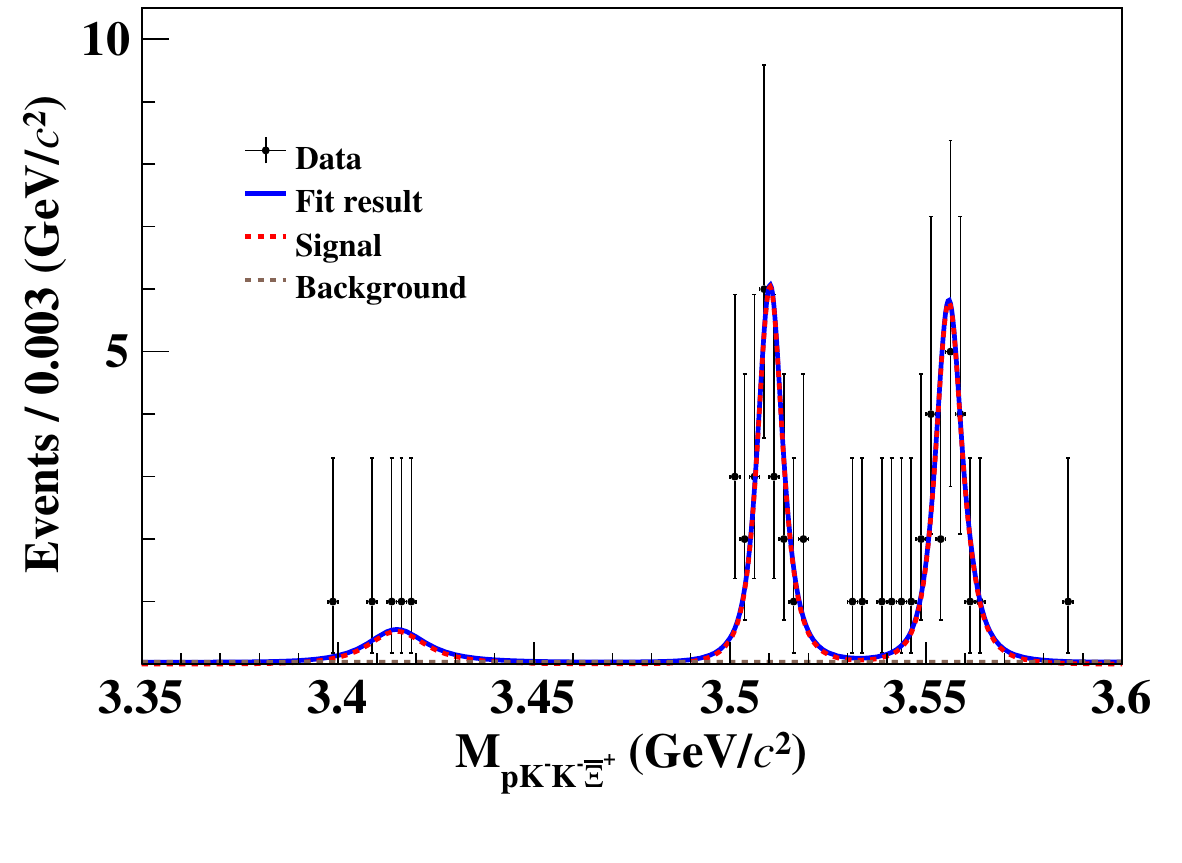}
	\vspace{-0.9cm}
	\caption{The unbinned maximum likelihood fit to the $M_{p K^- K^- \bar \Xi^+ }$ distribution of the accepted candidates of
	 $\chi_{cJ}\to p K^- K^- \bar \Xi^+$. The dots with error bars are data.
	The blue solid curve shows the full fit, and the red and brown dashed curves represent the signal and background shapes, respectively.
	}
	\label{tab:fit}
\end{figure}

 The invariant mass $M_{pK^{-}K^{-}\bar{\Xi}^{+}}$ distribution is used to discriminate signal from background. Since there are no sideband events from data, the signal yields are extracted from unbinned maximum likelihood fits to the $M_{pK^{-}K^{-}\bar{\Xi}^{+}}$ distributions
of events in the $\bar{\Xi}^{+}$ signal region for the three $\chi_{cJ}$ decays, as shown in Fig.~\ref{tab:fit}.
The $\chi_{cJ}$ signal shapes are modeled according to the MC simulation,
while the background shape is parameterized by a first-order polynomial.
The net signal yields \(N_{\text{net}}\) for
\(\chi_{c0}\), \(\chi_{c1}\), and \(\chi_{c2}\) are determined to be \(4.5_{-2.0}^{+2.6}\), \(22.4_{-4.6}^{+5.2}\), and \(23.0_{-4.7}^{+5.4}\), respectively.
The statistical significances of the \(\chi_{c0}\), \(\chi_{c1}\), and \(\chi_{c2}\) signals are \(2.8\sigma\), \(9.0\sigma\), and \(9.0\sigma\), determined from $\Delta(\mathrm{ln}\mathcal{L})=\mathrm{ln}\mathcal{L}_\mathrm{max}-\mathrm{ln}\mathcal{L}_0$ with $\Delta{ndf}=1$, where $\mathcal{L}_\mathrm{max}$ and $\mathcal{L}_\mathrm{0}$ are the maximum likelihoods with and without the signal component in the fit, respectively. The $\Delta{ndf}$  is the variation of number of degrees of freedom.

\begin{figure*}[htbp]
\centering
\scalebox{1.4}{
	\hspace{-0.15cm}
	\includegraphics[width=4.2cm]{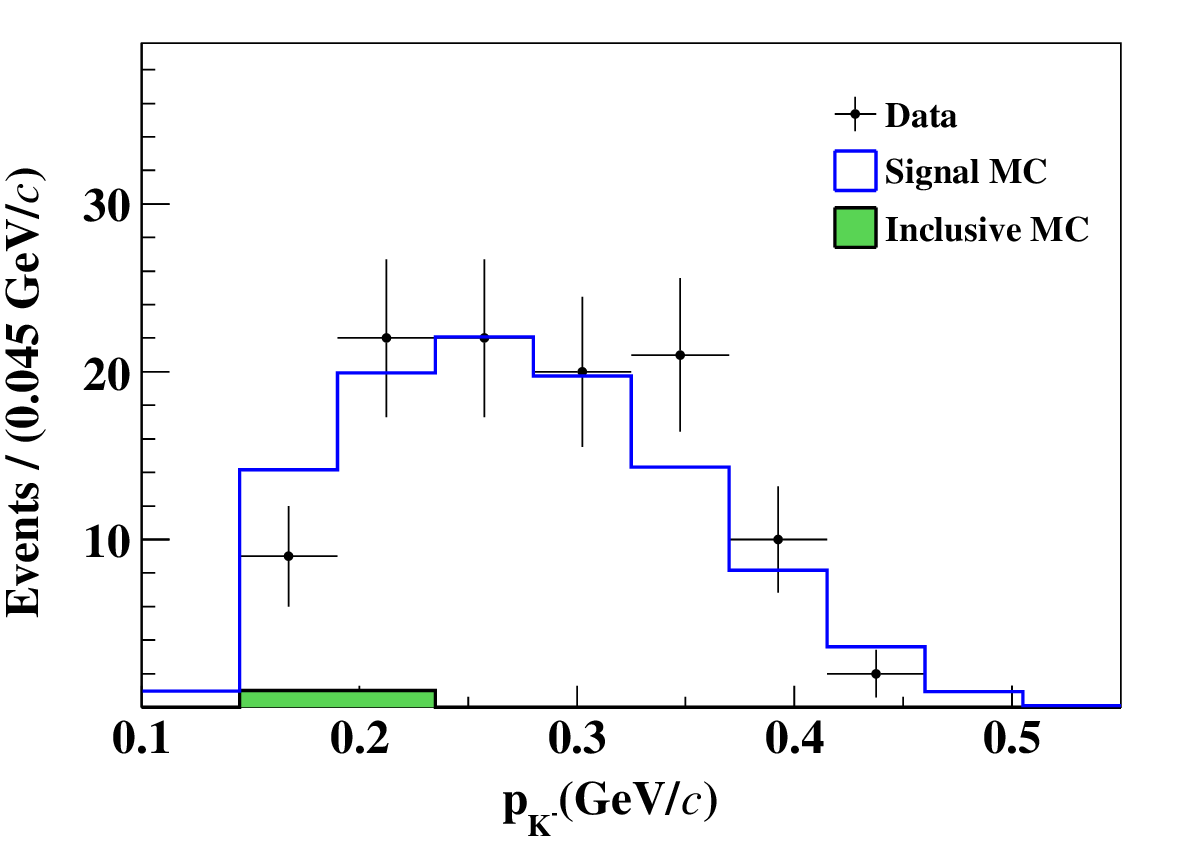}
	\includegraphics[width=4.2cm]{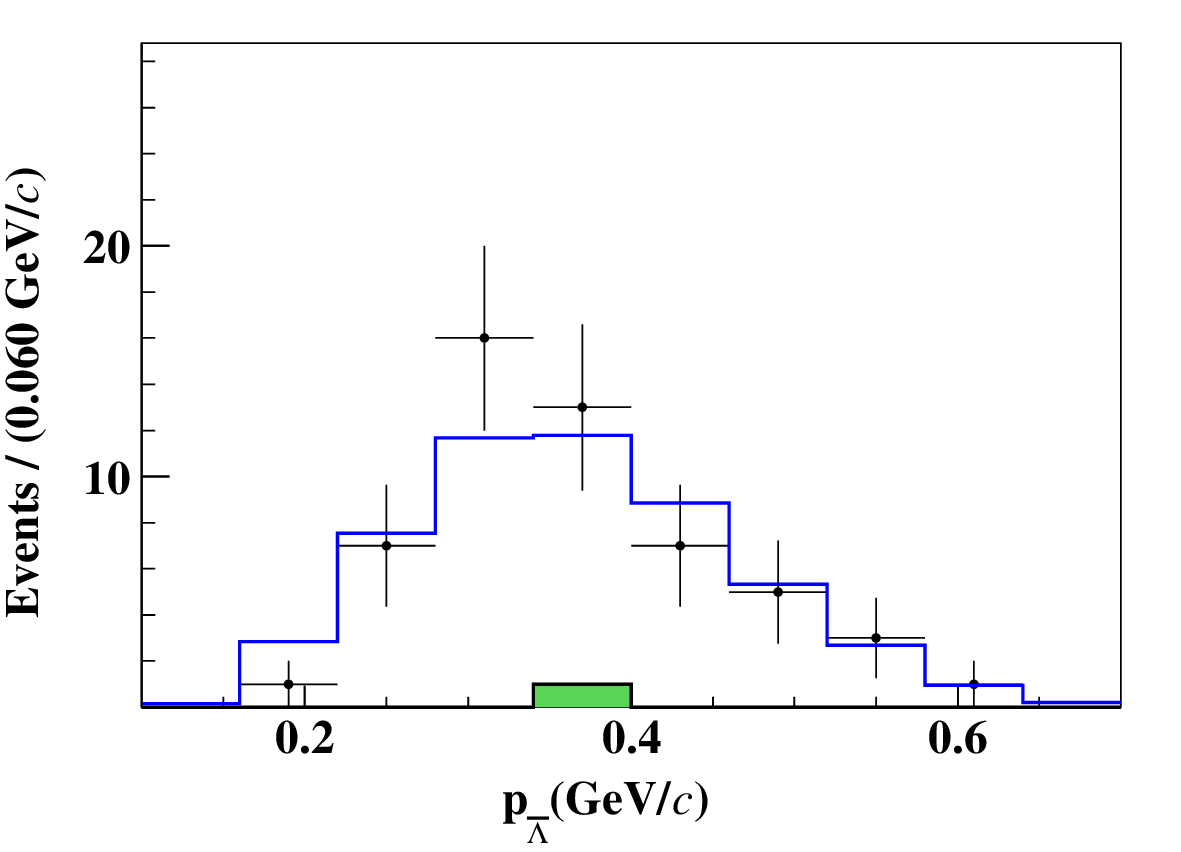}
	\includegraphics[width=4.2cm]{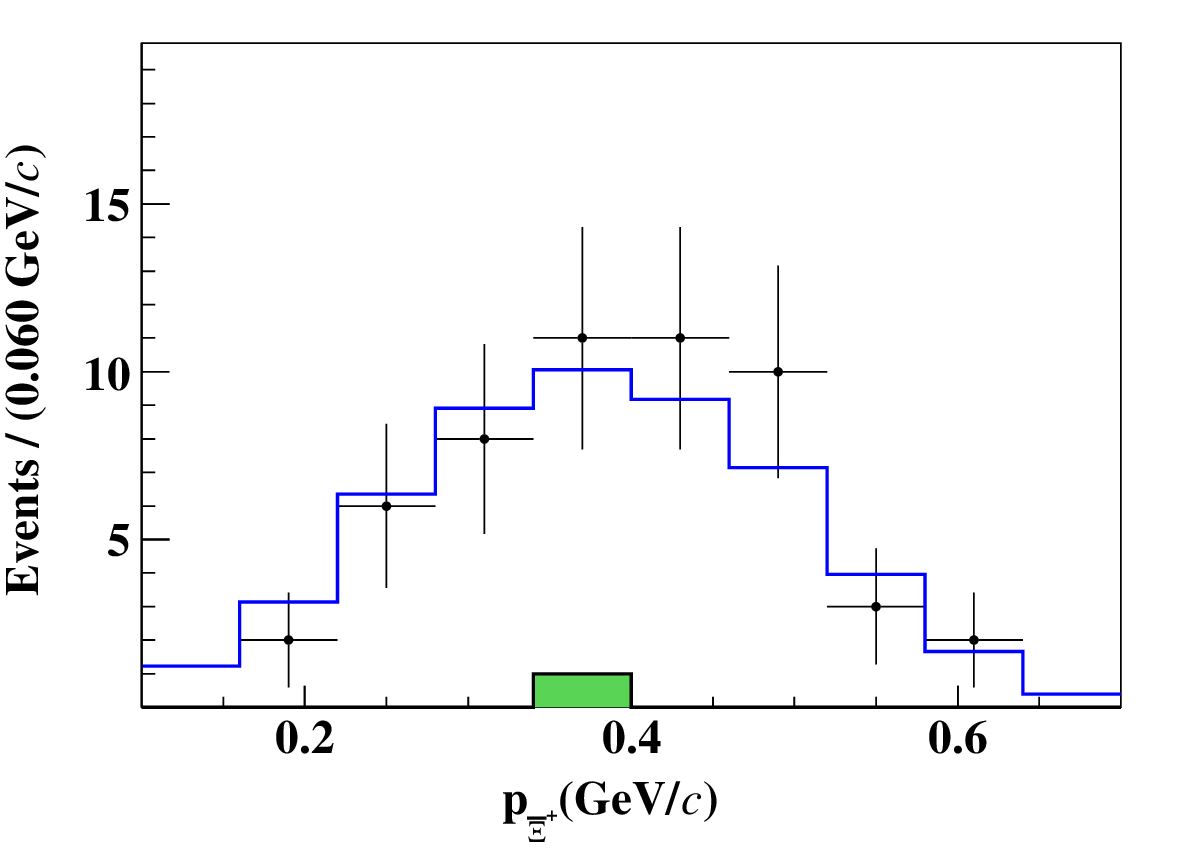}

}
\scalebox{1.4}{
	\hspace{-0.15cm}
	\includegraphics[width=4.2cm]{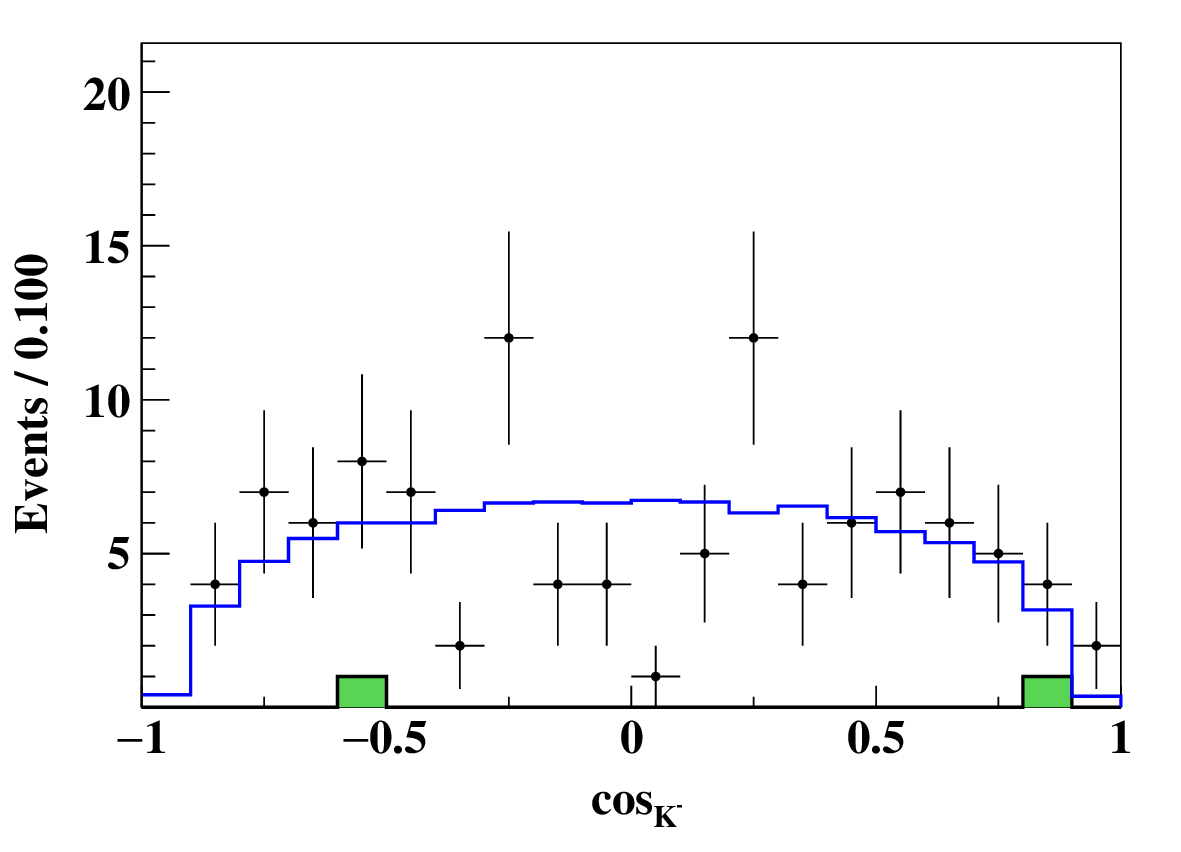}
	\includegraphics[width=4.2cm]{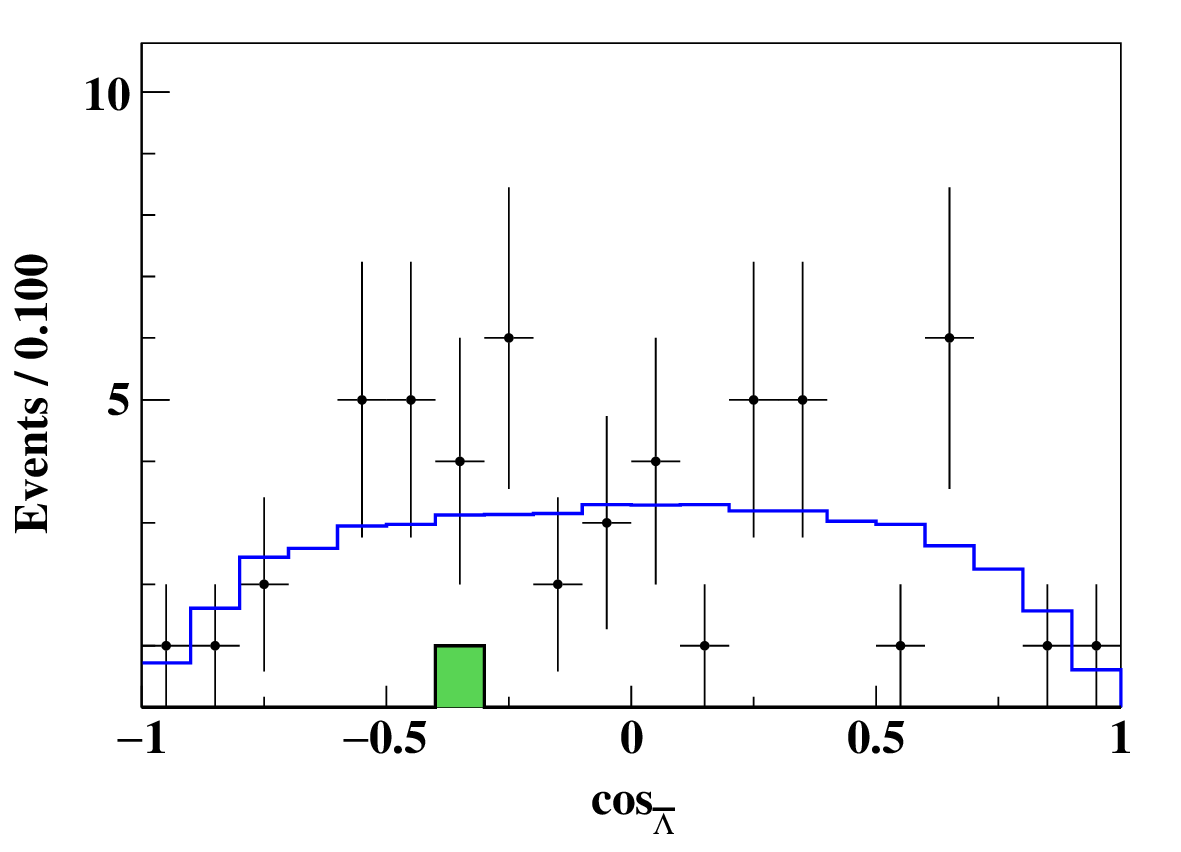}
	\includegraphics[width=4.2cm]{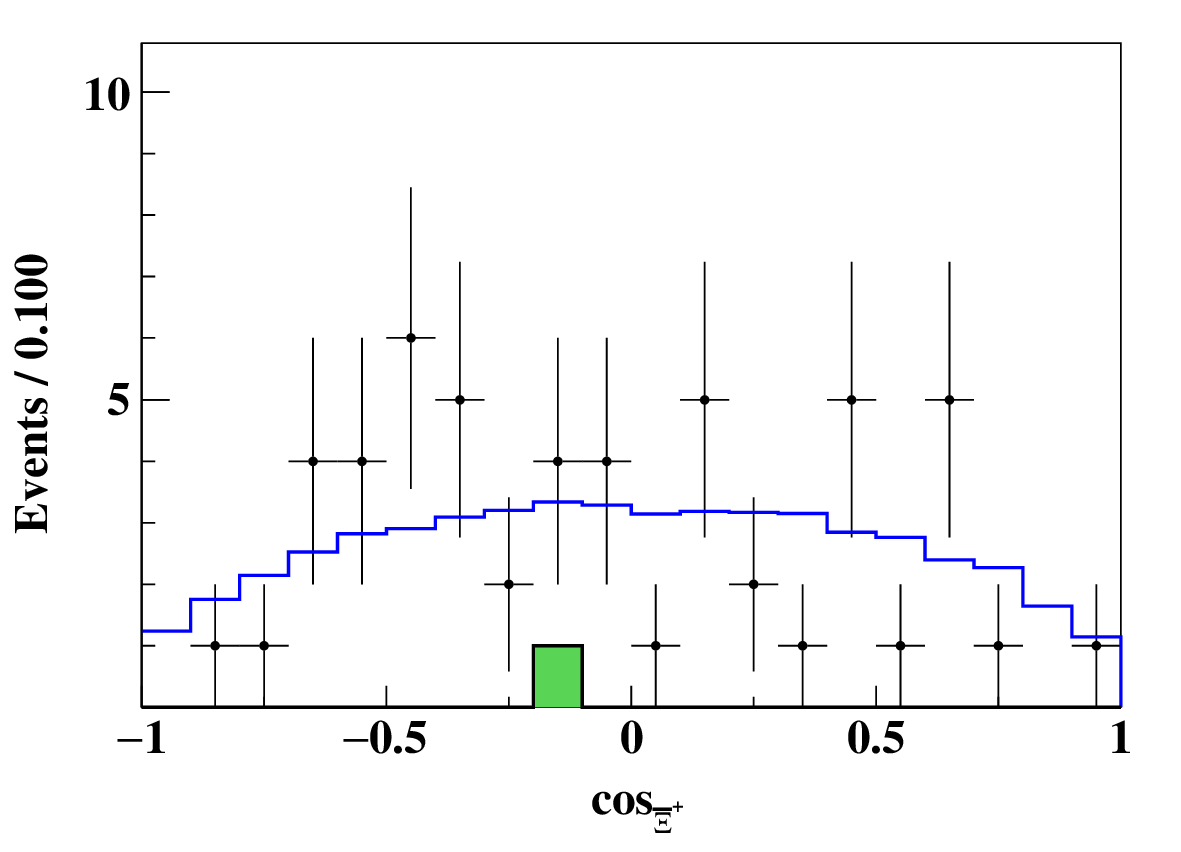}

}
\scalebox{1.4}{
	\hspace{-0.15cm}
	\includegraphics[width=4.2cm]{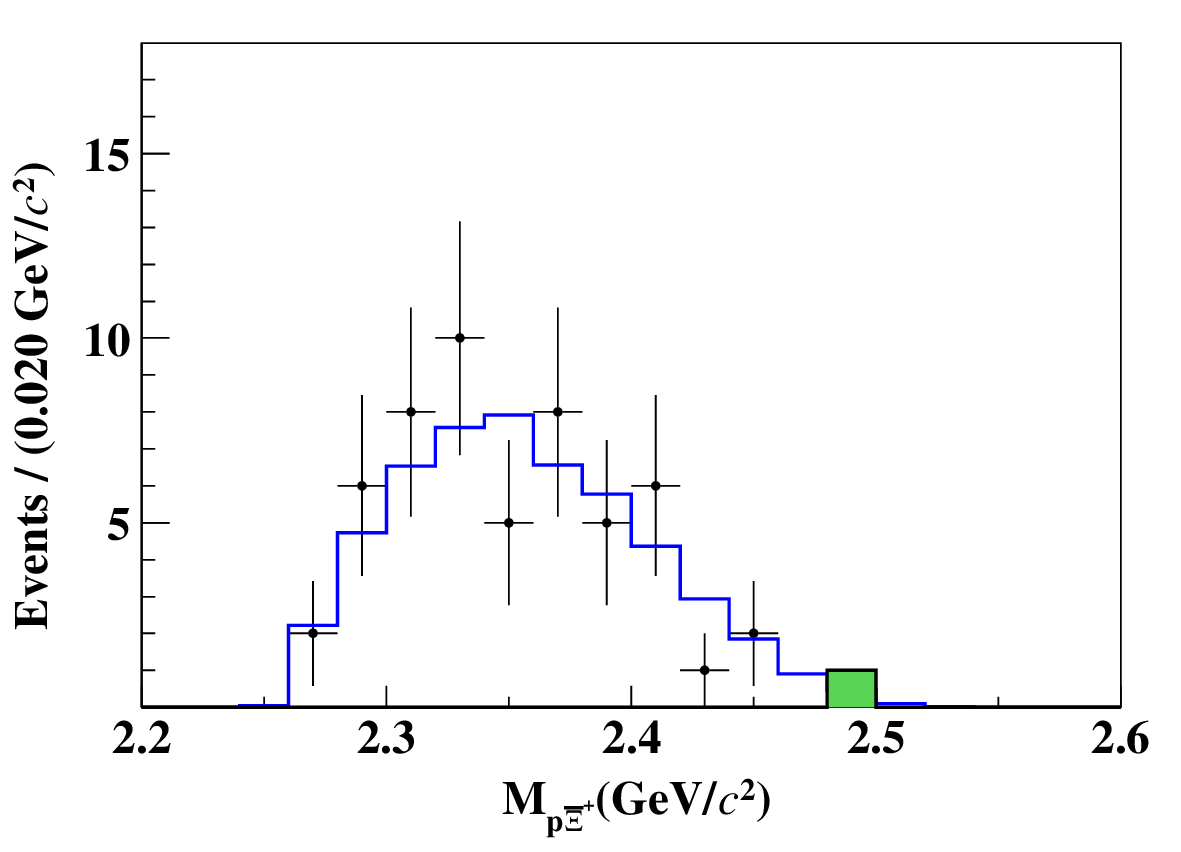}
    \includegraphics[width=4.2cm]{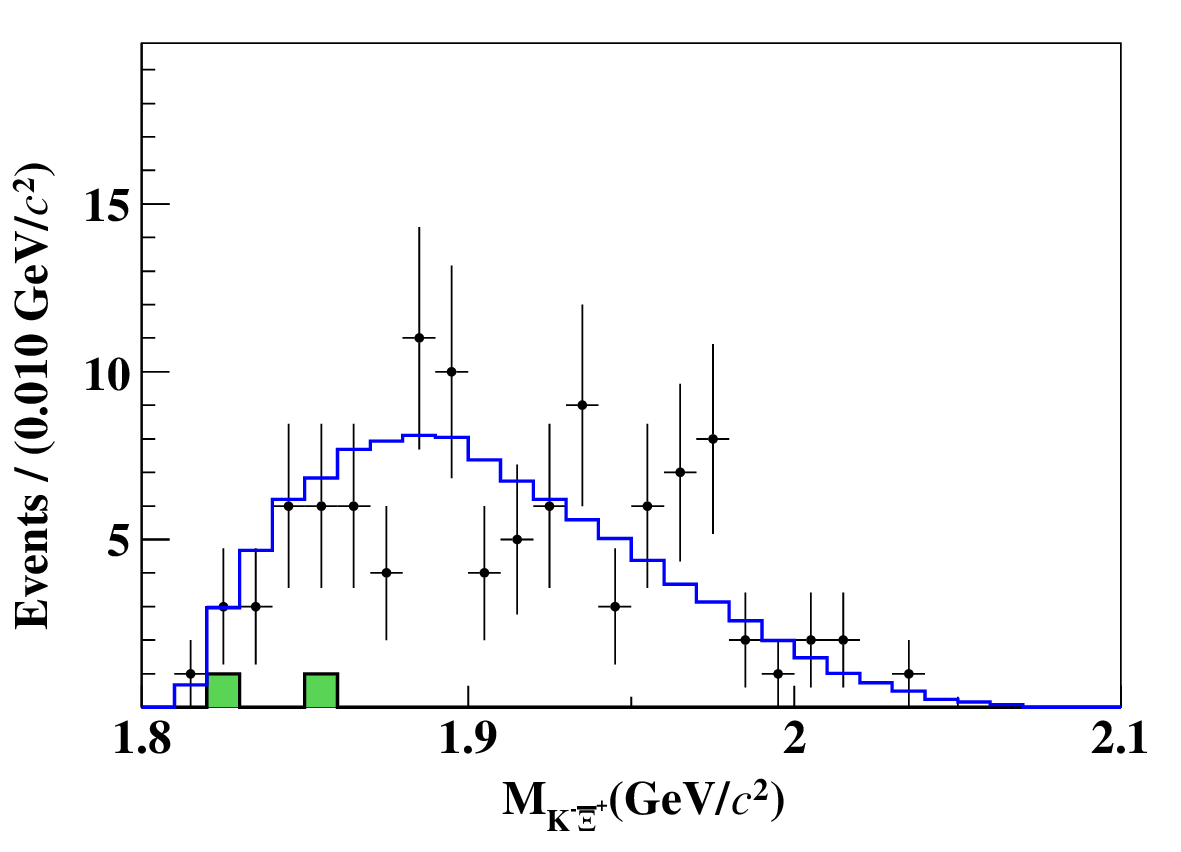}
	\includegraphics[width=4.2cm]{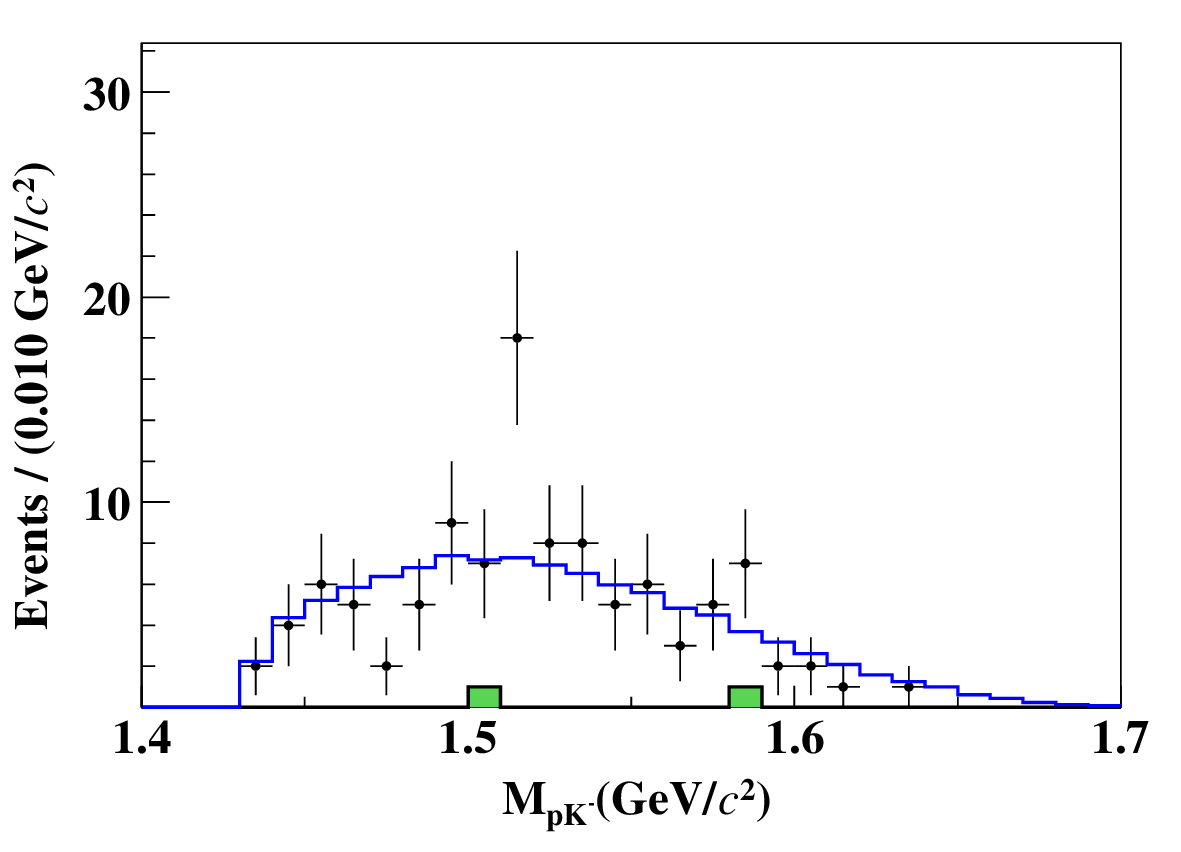}
}
	\caption{The comparisons of the momentum and polar angle cosine ditributions of daughter particles and two-body invariant mass distributions, for accepted candidate events for $\chi_{cJ} \to p K^- K^- \bar \Xi^+$.
	 The black dots with error bars represent data, blue histograms represent the signal MC sample, and green histograms represent simulated background events from the inclusive MC sample. The distributions involving $K^-$ have two entries per event, and all events are required to be within the $\bar \Xi^+$ signal region.}
	\label{tab:dataMC}
\end{figure*}

Figure~\ref {tab:dataMC} shows the momentum and polar angle cosine distributions of the daughter particles, along with the two-body invariant mass distributions.
The good agreement between data and MC simulation in both the momentum and angular spectra validates the reliability of the detection efficiencies.

\section{SYSTEMATIC UNCERTAINTY}
\label{sec:systematics}
The systematic uncertainties associated with the BF measurements arise from several sources, as detailed in Table~\ref{tab:systematics}.
Each contribution is evaluated and discussed as follows.

\begin{table*}[htbp]
	\centering
  	\caption{Relative systematic uncertainties (in unit of \%)  in the measurements of the BFs of $\chi_{cJ}\to p K^- K^- \bar \Xi^+$.
	The upper scripts with $^*$ apply for the determination of the upper limit for the $\chi_{c0}$ decay.}
	\vspace{4pt}
	\renewcommand{\arraystretch}{1.4}
  \begin{tabular}{lccc}
  	\hline
  	\hline
  	Source & $\chi_{c0}$ & $\chi_{c1}$ & $\chi_{c2}$  \\ \hline
  	$N_{\psi(3686)}$  & $ 0.5$ & $ 0.5$ & $ 0.5$ \\
  	Tracking &        $ 4.0$ & $ 4.0$ & $ 4.0$ \\
	PID     &         $ 4.0$ & $ 4.0$ & $ 4.0$ \\
  	 Photon~detection & $ 1.0$ & $ 1.0$ & $ 1.0$\\
	$\Lambda$ reconstruction & $ 2.5$ & $ 2.5$ & $ 2.5$\\
	$\Xi$ mass window & $ 0.5$ & $ 0.5$ & $ 0.5$\\
  	4C ~kinematic ~fit & $ 2.9$ & $ 2.0$ &$ 1.5$ \\
  	Invariant mass fit & $ 3.1$ & $ 1.0$ & $ 1.3$ \\
	 MC ~statistics & $2.2$ & $ 1.2$ & $ 1.0$\\
	 MC ~model & $\rm{negligible}$ & $ 4.8$ &$ 1.9$ \\
	$\mathcal B_{\psi(3686) \to \gamma \chi_{cJ}}$   &$2.0$ & $2.5$ &$2.1$ \\
	$\mathcal B_{\bar \Xi^+ \to \bar \Lambda \pi^+}$ & $0.1$ & $ 0.1$ & $ 0.1$ \\
	$\mathcal B_{\bar \Lambda \to \bar p \pi^+}$ & $0.8$ & $ 0.8$ & $ 0.8$ \\ \hline
	Total for ${\mathcal B}_{\psi(3686)\to \gamma \chi_{cJ}}\cdot {\mathcal B}_{\chi_{cJ}\to p K^- K^- \bar
\Xi^+}$~ & $8.0$  & $8.4$ & $7.0$ \\  \hline
  	Total for ${\mathcal B}_{\psi(3686)\to \gamma \chi_{cJ}}\cdot {\mathcal B}_{\chi_{cJ}\to p K^- K^- \bar
\Xi^+}$ $\rm (Without~uncertainty~due~to~invariant~mass~fit)^*$ & $7.3$  & $-$ & $-$ \\  \hline
	Total for ${\mathcal B}_{\chi_{cJ}\to p K^- K^- \bar \Xi^{+}}$ & $ 8.2$ & $8.7$ & $7.3$\\ \hline
	Total for ${\mathcal B}_{\chi_{cJ}\to p K^- K^- \bar \Xi^{+}}$ $\rm (Without~uncertainty~due~to~invariant~mass~fit)^*$& $7.
6$ & $-$ & $-$\\
  	\hline
  	\hline
  \end{tabular}
  \label{tab:systematics}
\end{table*}

The total number of $\psi(3686)$ events is determined to be $N_{\psi(3686)}=(2.712\pm0.014)\times10^9$ using the inclusive hadronic events as described in Ref.~\cite{psip-num-inc}, with a corresponding uncertainty of 0.5\%.

The tracking efficiencies of the proton, kaon and pion are studied using control samples of $J/\psi\to p\bar p\pi^+\pi^-$ and $J/\psi\to K^*(892)K$~\cite{ref::pid}, and the differences in tracking efficiency between data and MC simulation are assigned as systematic uncertainties of $1.0\%$ per track. The proton, kaon and pion PID uncertainties are also assigned as $1.0\%$ per track using the same control samples.

The photon detection uncertainty is assigned as 1.0\% per photon based on studies of the control sample $J/\psi\to\pi^+\pi^-\pi^0$~\cite{ref::gamma-recon}.

The systematic uncertainty of the $\bar\Lambda$ reconstruction is evaluated using control samples of $J/\psi\to pK^-\Lambda$ and $J/\psi\to \Lambda\bar \Lambda$, accounting for the tracking efficiencies of the $\bar p\pi^+$ pair, the decay length requirement, the $\bar \Lambda$ mass window,
the vertex fit, and the second vertex fit.
The resulting systematic uncertainty is estimated as $2.5\%$ per $\bar \Lambda$~\cite{ref::lambdaerror}.

The uncertainty associated with $\Xi$ mass window is evaluated using the control sample $\psi(3686) \to pK^-K^-\bar \Xi^+$,
with the discrepancy in acceptance efficiency between data and MC simulation taken as the systematic uncertainty.

The systematic uncertainty from the 4C kinematic fit is evaluated by comparing signal efficiencies with and without the helix parameter correction,
with correction factors taken from the previous study~\cite{ref::helixp}.
The resulting differences in signal efficiency are assigned as the systematic uncertainties,
which are 2.9\%, 2.0\% and 1.5\% for $\chi_{c0,1,2}\to p K^- K^- \bar \Xi^+$, respectively.

The systematic uncertainties of the fit to the $M_{p K^- K^- \bar \Xi^{+}}$ distribution are evaluated from three sources.
The uncertainty from background shape is assessed by replacing the first-order polynomial function with no background component.
The signal shape uncertainty is evaluated by varying the Gaussian kernel widths of the RooKeysPdf, corresponding to different levels of smoothness.
The uncertainty due to the fit range is assessed using the Barlow test method~\cite{Barlow:2002yb}, by repeating the fit with alternative ranges within [3.350,3.600]~GeV/$c^2$ in steps of
0.002~GeV/$c^2$. For each alternative, the deviation from the nominal fit is define as
\begin{linenomath*}
\begin{equation}
\delta = \frac{N'_{\rm net}-N_{\rm net}}{\sigma'^2-\sigma^2},
\end{equation}
\end{linenomath*}
where $N_{\rm Net}$ is the fitted signal yield and $\sigma$ is its statistical uncertainty. If $\delta<2.0$, the associated systematic uncertainty is considered negligble; otherwise, the largest relative difference is taken as the systematic uncertainty.
Combining these three contributions in quadrature gives total systematic uncertainties of 3.1\%, 1.0\%, and 1.3\% for $\chi_{c0,1,2} \to p K^- K^- \bar \Xi^{+}$, respectively.

The uncertainties due to limited signal MC statistics are estimated as
$\frac{1}{\sqrt{N}}\sqrt{\frac{1-\epsilon}{\epsilon}}$,
where $\epsilon$ is the detection efficiency and $N$
is the total number of signal MC events, yielding uncertainties of 2.2\%, 1.2\%, and 1.0\% for $\chi_{c0,1,2}\to p K^- K^- \bar \Xi^+$, respectively.

The systematic uncertainty from the MC model is evaluated by examining the effect of including the intermediate state $\psi(3686) \to \Lambda(1520) K^- \bar{\Xi}^{+}$ on the signal efficiency. The $M_{pK^-}$ distribution of accepted candidates in data is fitted with a signal shape modeled from MC simulation and a first-order polynomial background.
No significant $\Lambda(1520)$ signal is observed in the $\chi_{c0}$ decay, while the $\Lambda(1520)$ fractions in
$\chi_{c1,2}\to p K^- K^- \bar{\Xi}^+$ are measured to be $(45\pm21)\%$ and $(43\pm21)\%$, respectively.
The corresponding changes in signal efficiency of 4.8\% and 1.9\% are assigned as the systematic uncertainties for $\chi_{c1,2}$, while this uncertainty is considered negligible for $\chi_{c0}$ due to the absence of evidence for
the $\Lambda(1520)$ intermediate state.

The BFs for the $\chi_{cJ}$, $\bar \Xi^{+}$, and $\bar \Lambda$ decays are taken from the PDG~\cite{ref::pdg2024}, with corresponding uncertainties of 2.0\%, 2.5\%, and 2.1\% for $\psi(3686)\to \gamma\chi_{c0,1,2}$, 0.1\% for $\bar \Xi^+ \to \bar \Lambda \pi^+$
and 0.8\% for $\bar \Lambda \to \bar p \pi^+$.

For each signal decay, the total systematic uncertainty is obtained by adding all individual contributions in quadrature.

\section{BRANCHING FRACTION AND UPPER LIMIT}
The BFs of $\chi_{cJ} \to p K^- K^- \bar \Xi^+ $ are calculated as
\begin{equation}
\begin{aligned}
\mathcal B_{\chi_{cJ} \to  p K^- K^- \bar \Xi^+} =\frac{N_{\rm net}}{N_{\psi(3686)} \cdot\mathcal B_{\rm sub} \cdot {\mathcal \epsilon} \cdot \mathcal B_{\psi(3686) \to \gamma \chi_{cJ}}},
\end{aligned}
\end{equation}
where $N_{\psi(3686)}$ is the total number of $\psi(3686)$ events in data sample, $\epsilon$ is the detection efficiency determined from MC simulation, and
$\mathcal B_{\rm sub}$ is the product of  $\mathcal B_{\bar \Xi^+ \to \bar \Lambda \pi^{+}}$ and $\mathcal B_{\bar \Lambda \to \bar p \pi^{+}}$,
taken from the PDG~\cite{ref::pdg2024}.
The resulting BFs are summarized in Table~\ref{tab:branching}.

\begin{table*}[htbp]
	\centering
	\caption{The quantities used for determining BFs of $\chi_{cJ}\to p K^- K^- \bar \Xi^+$, where the first uncertainties are statistical and the second systematic. The numbers in brackets correspond to the upper limits at the 90\% confidence level.}
\vspace{4pt}
\renewcommand{\arraystretch}{1.4}
	\begin{tabular}{cccc}
		\hline
		\hline                                          & $\chi_{c 0}$           & $\chi_{c 1}$            & $\chi_{c 2}$    \\
		\hline
		$N_{\text{net}}$                    & $4.5_{-2.0}^{+2.6}$ ~($<8.9$)   & $22.4_{{-4.6}}^{+5.2}$    &
$23.0_{-4.7}^{+5.4}$ ~    \\
		$\epsilon$~(\%)                                 & 0.41                   & 1.37                    & 1.86    \\
		Significance ($\sigma$)                        & 2.8                   & 9.0                    & 9.0    \\
		$\mathcal{B}_{\psi(3686) \to \gamma \chi_{cJ}} \cdot  \mathcal{B}_{\chi_{cJ} \to p K^{-} K^{-} \bar{\Xi}^{+}+c.c.} (\times10^{-6}) $            &
$0.63_{-0.28}^{+0.36}\pm0.04(<1.24)$     & $0.94_{-0.19}^{+0.22}\pm0.07$        &$0.71_{-0.15}^{+0.17}\pm0.04$ \\
		$\mathcal{B}_{\chi_{cJ} \to p K^{-} K^{-} \bar{\Xi}^{+}+c.c.} (\times10^{-6}) $            & $6.4_{-2.9}^{+3.7}\pm0.5(<12.7)$     &
$9.6^{+2.2}_{-2.0}\pm0.8$        &$7.5^{+1.8}_{-1.5}\pm0.5$ \\
		\hline \hline
	\end{tabular}
	\label{tab:branching}
\end{table*}

Since the significance of $\chi_{c0} \to pK^-K^- \bar \Xi^+$ is only 2.8$\sigma$,
an upper limit on its BF is set at the 90\% confidence level.
The systematic uncertainties are incorporated following the same methodology used for the BF measurements, with the distinction between additive and multiplicative contributions.
The uncertainty from the invariant mass fit is treated as additive uncertainty, meaning it is independentof the measured BF central value, while all other uncertainties listed in Table~\ref{tab:systematics} are treated as multiplicative uncertainties, meaning they scale proportionally with the branching fraction.

For multiplicative uncertainties, the systematic contributions marked with $^*$ in Table~\ref{tab:systematics} are incorporated in the determination of the upper limit on the $\chi_{c0}\to p K^- K^- \bar \Xi^{+}$ decay.
For additive uncertainties, two effects are considered: the signal shape uncertainty, evaluated by varying the smoothness of the signal shape, and the background shape uncertainty, assessed by replacing the first-order polynomial with no background component. The most conservative upper limit obtained across all additive variations is taken as the final result.

\begin{figure}[htbp]
	\centering
	\scalebox{1.18}{
	\hspace{-0.31cm}
	\includegraphics[width=8.0cm]{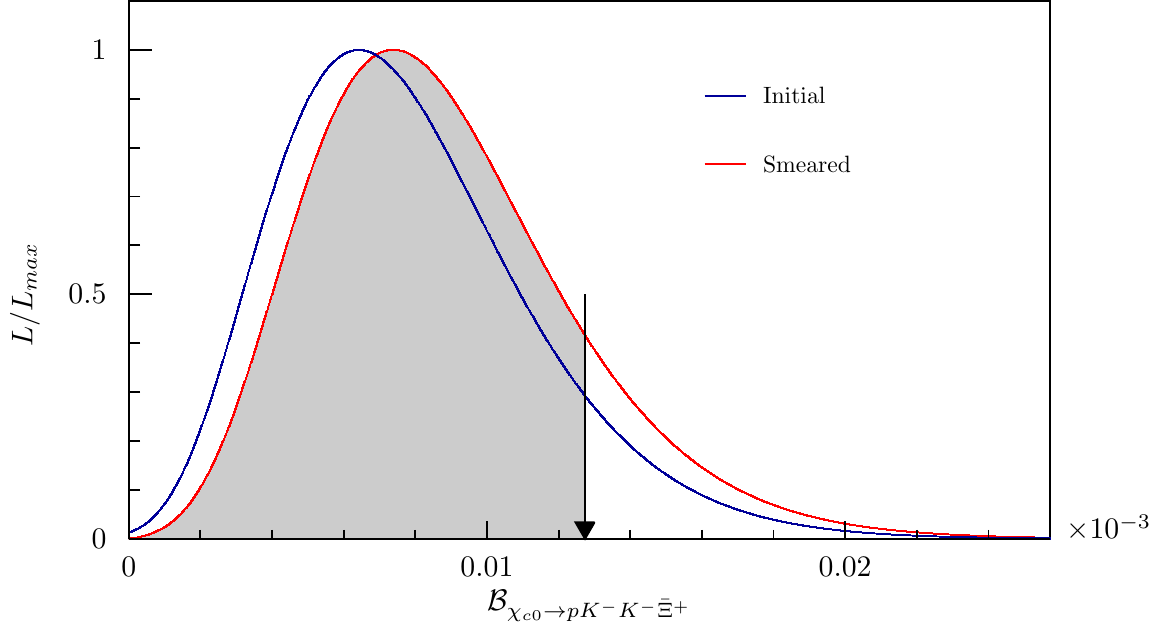}
      }
      	\caption{The normalized likelihood distribution as a function of the BF of $\chi_{c0}\to p K^- K^- \bar{\Xi}^{+}$. The red and blue solid curves correspond to the results with and without systematic uncertainties incorporated, respectively. The black arrow indicates the upper limits at the 90\% confidence level.}

	\label{fig5}
\end{figure}

To incorporate the multiplicative systematic uncertainties in the upper limit calculation, the likelihood distribution is smeared by a Gaussian function of zero mean and width equal to the multiplicative uncertainty,
$\sigma_\varepsilon$~\cite{scan},
\begin{equation}
L(\mathcal{\mathcal{B}})\propto\int_0^1L\left(\mathcal{\mathcal{B}}\cdot\frac{\varepsilon}{\varepsilon_0}\right)\cdot e^{-\frac{\left(\frac{\varepsilon}{\varepsilon_0}-1\right)^2}
{2\sigma_\varepsilon^2}}d\varepsilon,
\end{equation}
where $\mathcal{B}$ is the true (expected) BF, $L(\mathcal{B})$ is the likelihood as a function of the assumed BF, and $\mathcal{B}\cdot\frac{\varepsilon}{\varepsilon_0}$ represents the possible measured values of the BF with associated likelihood $L(\mathcal{B}\cdot\frac{\varepsilon}{\varepsilon_0})$. Here, $\varepsilon_0$ is the detection efficiency from MC simulation, and $\varepsilon$ is the true (expected) efficiency, treated as a random variable smeared around $\varepsilon_0$ according to a Gaussian to account for systematic uncertainties in the MC estimation.
The normalized likelihood as a function of the BF of $\chi_{c0}\to p K^- K^- \bar{\Xi}^{+}$ is shown in Fig.~\ref{fig5}.

\section{Summary}

Using  $(2.712\pm0.014)\times10^9$ $\psi(3686)$ events collected with the BESIII detector,
we report the first measurements of branching fractions of $\chi_{cJ}\to p K^- K^- \bar{\Xi}^{+}~(J=0,1,2)$, with the results summarized in Table~\ref{tab:branching}.
These represent the first observation of $\chi_{cJ}$ decays into final states containing three strange particles in the $\chi_{cJ}\to BB'KK$ system, providing a unique input for understanding the decay mechanisms of charmonium states.
The substantially larger $\psi(3686)$ data samples expected from the future super tau-charm factory~\cite{ref::fac} will enable more detailed studies of the intermediate states contributing to these decays.

\section{Acknowledgement}
The BESIII Collaboration thanks the staff of BEPCII (\href{https://lssf.cas.cn/sszs/zyyj/bjzfdzdzj/}{https://cstr.cn/31109.02.BEPC}) and 
the IHEP computing center for their strong support. This work is supported in part by National Key R\&D Program 
of China under Contracts Nos. 2025YFA1613900,
2023YFA1606000, 2023YFA1606704; National Natural Science Foundation of China (NSFC) under Contracts Nos. 11635010, 11935015, 11935016, 
11935018, 12025502, 12035009, 12035013, 
12061131003, 12192260, 12192261, 12192262, 
12192263, 12192264, 12192265, 12221005, 12225509, 12235017, 12342502, 12361141819, 12535005; the Chinese Academy of Sciences (CAS) 
Large-Scale Scientific Facility Program; the Strategic Priority 
Research Program of Chinese Academy of 
Sciences under Contract No. XDA0480600; CAS under Contract No. YSBR-101; 100 Talents Program of CAS; The Institute of Nuclear and 
Particle Physics (INPAC) and Shanghai Key Laboratory for Particle 
Physics and Cosmology; Agencia Nacional de Investigación y Desarrollo de Chile (ANID), Chile under Contract No. ANID CCTVal CIA250027; 
ERC under Contract No. 758462; German Research Foundation DFG under Contract No. FOR5327; Istituto 
Nazionale di Fisica Nucleare, Italy; Knut 
and Alice Wallenberg Foundation under Contracts Nos. 2021.0174, 2021.0299, 2023.0315; Ministry of Development of Turkey under Contract 
No. DPT2006K-120470; National Research Foundation of Korea under 
Contract No. RS-2026-25486791; 
National Science and Technology fund of Mongolia; Polish National Science Centre under Contract No. 2024/53/B/ST2/00975; STFC (United 
Kingdom); Swedish Research Council under Contract No. 2019.04595; 
U. S. Department of Energy under 
Contract No. DE-FG02-05ER41374.

\end{document}